\documentclass[11pt]{article}
\usepackage{amsmath}
\usepackage{latexsym}
\usepackage{graphicx}
\usepackage{bbm}
\usepackage{color}
\usepackage{euscript}
\usepackage{amsfonts}
\usepackage{cite}

\makeatletter
\@addtoreset{equation}{section}
\makeatother

\def\beq{\begin{equation}}
\def\eeq{\end{equation}}
\def\be{\begin{equation}}
\def\ee{\end{equation}}

\def\bea{\begin{eqnarray}}
\def\eea{\end{eqnarray}}
\def\ft#1#2{{\textstyle{\frac{\scriptstyle #1}{\scriptstyle #2}}}}
\def\fft#1#2{\frac{#1}{#2}}

\let\ep=\epsilon

\let\m=\mu 
\let\n=\nu

\let\t=\tau

\let\bm=\bibitem

\newcommand{\tr}{{\rm tr} }

\def\R{{\mathbbm R}}

\def\Z{{\mathbbm Z}} 

\def\cK{{\cal K}}
\def\cG{{\cal G}}
\def\cH{{\cal H}}
\def\cV{{\cal V}}

\def\bG{{\bf G}}
\def\bH{{\bf H}}
\def\bK{{\bf K}}
\def\del{{\partial}}
\def\bm{\bibitem}
\def\sst#1{{\scriptscriptstyle #1}}
\def\n{{\sst (n)}}
\def\m{{\sst (m)}}

\def\nn{\nonumber}
\def\td{\tilde}

\def\ie{{\it i.e.\ }} 

\newcount\hour \newcount\minute
\hour=\time  \divide \hour by 60
\minute=\time
\loop \ifnum \minute > 59 \advance \minute by -60 \repeat
\def\nowtwelve{\ifnum \hour<13 \number\hour:
                      \ifnum \minute<10 0\fi
                      \number\minute
                      \ifnum \hour<12 \ A.M.\else \ P.M.\fi
	 \else \advance \hour by -12 \number\hour:
                      \ifnum \minute<10 0\fi
                      \number\minute \ P.M.\fi}
\def\nowtwentyfour{\ifnum \hour<10 0\fi
		\number\hour:
         	\ifnum \minute<10 0\fi
         	\number\minute}

\begin{document}

\voffset=0.3truein
\hfuzz=100pt

\title{Infinite-Dimensional Symmetries of Two-Dimensional Coset Models
 Coupled to Gravity}
\author{\Large H. L\"u,$^1$ Malcolm J. Perry$^2$ and
C.N. Pope$^{1,2}$
\\ \\ \\
${}^1$ George P. \& Cynthia W. Mitchell Institute for Fundamental Physics,\\
         Texas A\&M University,
         College Station,
         TX 77843-4242,
         USA.\\ \\
${}^2$ DAMTP, Centre for Mathematical Sciences,\\
         University of Cambridge,
         Wilberforce Road,
         Cambridge CB3 0WA,
         England.\\ \\ \\  \\} 

\date{\empty}
\maketitle

\includegraphics[scale = 0.05, bb= 0 -6800 100 -8600 ]{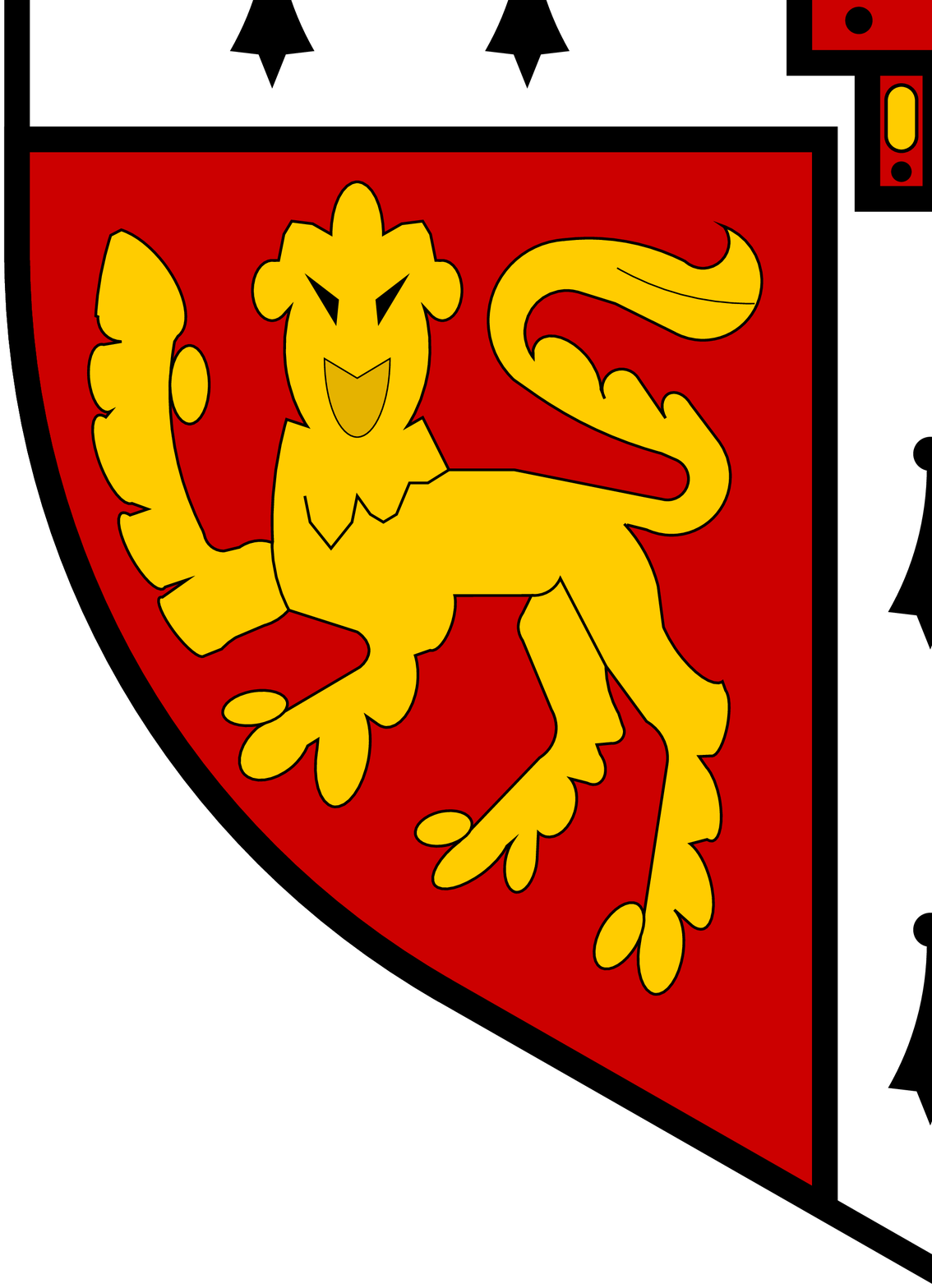}
\includegraphics[scale = 0.17, bb= -1900 -2000 0 -1800 ]{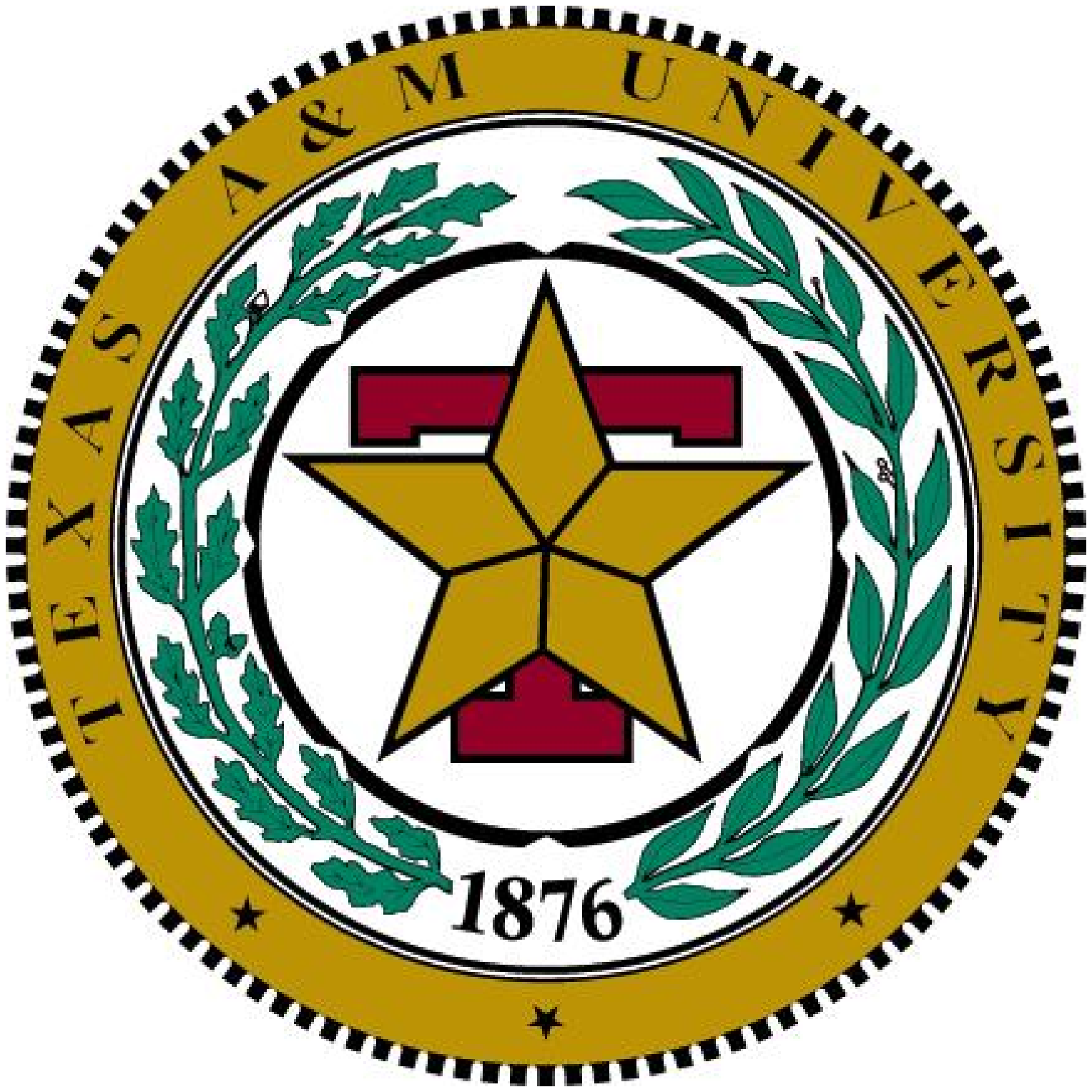}

\vskip -5.5truein

\centerline{DAMTP-2007-115\hskip 1.5truein MIFP-07-31}
\bigskip\bigskip
\centerline{\bf{ArXiv:0712.0615}}

\vskip 3.5truein


\begin{abstract} 

  In an earlier paper we studied the infinite-dimensional symmetries
of symmetric-space sigma models (SSMs) in a flat two-dimensional spacetime.
Here, we extend our investigation to the case of two-dimensional SSMs
coupled to gravity.  These theories arise from the toroidal reduction
of higher-dimensional gravity and supergravities to two dimensions.
We construct explicit expressions for the symmetry transformations under
the affine Kac-Moody extension $\hat\cG$ that arises when starting from
a $\cG/\cH$ coset model.  We also construct further explicit symmetry
transformations that correspond to the modes $L_n$ of a Virasoro subalgebra
with $n\ge -1$.

\end{abstract}

\thispagestyle{empty}

\pagebreak
\voffset=0pt
\setcounter{page}{1}

\tableofcontents

\addtocontents{toc}{\protect\setcounter{tocdepth}{2}}

\def\V{\mathcal{V}}
\def\hV{\hat{\mathcal{V}}}

\newpage

\section{Introduction}

   The study of supergravity theories, and their symmetries, have played
a very important r\^ole in uncovering the underlying structures of
string theory.  Especially significant are the U-duality symmetries of
the string, which have their origin in the classical global symmetries
exhibited by eleven-dimensional supergravity and type IIA and IIB supergravity
after toroidal dimensional reduction.  For example, if one reduces
eleven-dimensional supergravity on an $n$-torus, for $n\le8$, the
resulting $D=(11-n)$-dimensional theory exhibits a global $E_{n}$ symmetry
\cite{crejul,cj2,cjlp}.  In the cases $n\ge3$ this symmetry arises in quite a
subtle way, involving an interplay between the original eleven-dimensional
metric and the 3-form potential.  The global symmetry can be understood 
most simply by first focusing attention on the scalar-field sector of the
dimensionally-reduced theory.  The scalars are described by a non-linear
sigma model, and in fact the cosets that arise are always symmetric spaces. 
The case of reduction to three dimensions is in many ways the most elegant,
because all of the bosonic fields in the theory (aside from the metric) 
are now scalars, and so one has just an $E_8/O(16)$ symmetric-space 
sigma model in a gravity background.

   In view of the large $E_8$ symmetry that one finds after reduction to
three dimensions, it is natural to push further and investigate the symmetries
after further reduction to two dimensions, and even beyond.  It turns out that
the analysis for a reduction to two dimensions is considerably more
complicated than the higher-dimensional ones.  There are two striking new
features that lead to this complexity.  The first is that, unlike a
reduction to $D\ge3$ dimensions, one can no longer use a reduction
scheme in which the metric is reduced from an Einstein-frame metric in the
higher dimension to an Einstein-frame metric in the lower dimension.  (In
the Einstein conformal frame, the Lagrangian for gravity itself takes the
form ${\cal L} \sim \sqrt{-g} R$, with no scalar conformal factor.)  The
inability to reach the Einstein conformal frame in two dimensions is
intimately connected to the fact that $\sqrt{-g} R$ is a conformal
invariant in two dimensions.   

   The second striking new feature is that an axionic
scalar field (\ie a scalar appearing everywhere covered by a derivative)
can be dualised to give another axionic scalar field in the
special case of two dimensions.  This has the remarkable consequence that
the global symmetry group actually becomes infinite in dimension.  This
was seen long ago by Geroch, in the context of four-dimensional gravity
reduced to two.   There are degrees of freedom
in two dimensions that are described by the sigma model $SL(2,\R)/O(2)$,
and under dualisation this yields another $SL(2,\R)/O(2)$ sigma model.
Geroch showed that the two associated global $SL(2,\R)$ symmetries do
not commute, and that if one takes repeated commutators of the two
sets of transformations, an infinite-dimensional algebra results \cite{geroch}.
The precise nature of this symmetry, now known as the {\it Geroch Group},
was not uncovered in \cite{geroch}.

   It is of considerable interest, therefore, to study the general
case of symmetric-space sigma model in a three-dimensional gravitational
background, after performing a further circle reduction to two dimensions.
(We shall use the acronym SSM to denote a symmetric-space sigma model.)
One arrives at a system in two dimensions that comprises an SSM coupled to 
gravity, together with an additional scalar field which can be thought
of as the Kaluza-Klein scalar for the reduction from three to two dimensions.

   In a previous paper \cite{lupepo}, we studied the global symmetries
of the simpler situation where one has an SSM in a purely flat two-dimensional
background.  Our goal in the present paper is to extend our analysis to the
full gravitationally-coupled system that arises when one reduces a 
gravity-coupled SSM from three to two dimensions.  

   There is quite a considerable earlier literature on the subject of 
the infinite
dimensional symmetries of two-dimensional symmetric-space sigma models,
both in the flat and the gravity-coupled cases (see, for example,
\cite{pol,lus,polus,kinchi,belzak,mais1,breitmais,dolan,devfair,yswu,jul,%
nicolai,nicjul},
some of which considers also principal chiral models).  In our previous 
work on the flat-space SSM case in
\cite{lupepo}, we found the work of Schwarz in \cite{schwarz1} to be
notably accessible, and we took that as our starting point.  Starting
from a two-dimensional flat-space SSM based on the coset $\cG/\cH$, 
Schwarz gave an explicit construction of symmetry transformations whose 
commutators generated a certain subalgebra $\hat\cG_H$ of the affine Kac-Moody 
extension $\hat\cG$ of $\cG$.  He also constructed additional symmetry 
transformations that formed a subalgebra of the centreless Virasoro algebra. 
We were then able to extend Schwarz's results, and obtain explicit
results exhibiting the full $\hat\cG$ Kac-Moody symmetry.

   In the present paper, we again take as our starting point the work
of Schwarz, who had extended his flat-space analysis to the gravitationally
coupled case in \cite{schwarz2}.  We develop Schwarz's construction by 
first improving on the removal of certain singularities in the transformation
rules, and then in addition we are able again to extend Schwarz's 
construction of explicit $\hat G_H$ transformations to the full set
of transformations for the entire Kac-Moody algebra $\hat\cG$.  Schwarz
did not succeed in generalising his construction of  Virasoro-type 
transformations to the gravitationally-couple case.  We succeed in doing
this too.  An essential feature now is that the Kaluza-Klein scalar coming
from the descent from three dimensions transforms under the Virasoro-type
symmetries, even though it is inert under the Kac-Moody transformations.

   We also discuss, in an appendix, how one can take a decoupling limit
of our results, to recover our earlier results in \cite{lupepo} on the
infinite-dimensional symmetries of two-dimensional sigma models in the
absence of gravity.

\section{Lax Equation and Infinite-Dimensional Symmetries}

\subsection{Reduction from three dimensions}

   We take as our starting point a symmetric-space non-linear sigma
model defined on the coset manifold $\bK=\bG/\bH$, and coupled to gravity 
in three spacetime dimensions.  The commutation relations
for the corresponding generators of the Lie algebra $\cG$ take the form
\be
[\cH,\cH]=\cH\,,\qquad [\cH,\cK]=\cK\,,\qquad
   [\cK,\cK]=\cH\,.\label{coset}
\ee
The condition that $\bK$ is a symmetric space is reflected in the absence of
$\cK$ generators on the right-hand side of the last commutation relation.
The symmetric-space algebra implies that there is an involution $\sharp$ 
under which
\be
\cK^\sharp = \cK\,,\qquad \cH^\sharp=-\cH\,.
\label{invol1}
\ee
In many cases, such as when $\bG=SL(n,\R)/O(n)$, the involution map is given
by Hermitean conjugation,
\be
\cK^\dagger=\cK\,,\qquad \cH^\dagger
 = -\cH\,.\label{invol2}
\ee
In some cases,
such as $\bG=E_{(8,8)}$, $\cH=O(16)$, the involution $\sharp$ is more involved. 

   The fields of
the sigma model $\bG/\bH$ may be parameterised by a coset representative
$\cV$, in terms of which we may define
\be
M=\cV^\sharp\, \cV\,,\qquad A= M^{-1} dM\,.\label{MAdef}
\ee
Under transformations 
\be
\cV\longrightarrow h\cV g\,,\label{gtrans}
\ee
where $g$ is a global element
in the group $\bG$ and $h$ is a 
local element in the denominator subgroup $\bH$, we have shall have
\be
M\longrightarrow g^\sharp M g\,,\qquad 
                   A\longrightarrow g^{-1} A g\,,\label{Mtrans}
\ee
since it follows from $\cH^\sharp =-\cH$ that $h^\sharp= h^{-1}$.
Henceforth, we shall consider for simplicity cases where the involution 
$\sharp$ is just Hermitean conjugation.  For the general case, all
occurrences of $\dagger$ should be replaced by $\sharp$.

   The Cartan-Maurer equation $d(M^{-1} dM)= -(M^{-1} dM)\wedge (M^{-1} dM)$
implies that the field strength for $A$ vanishes:
\be
F\equiv dA + A\wedge A=0\,.\label{cartmau0}
\ee
The Lagrangian for the three-dimensional model may be written as 
\be
{\cal L}_3 = \sqrt{-\hat g} \, 
\big(\hat R - \ft14\hat g^{MN}\, \tr(A_M A_N)\big)\,,
\label{d3lag}
\ee
where $\hat g_{MN}$ is the three-dimensional spacetime metric tensor.  The
sigma-model equations of motion are therefore given by
\be
\hat\nabla^M A_M=0\,.
\ee

   We now assume that the metric and sigma-model fields are all 
independent of one of the coordinates, which we shall denote by $z$.  We
may reduce the metric according to the Kaluza-Klein ansatz
\be
d\hat s_3^2 = e^\psi ds_2^2 + \rho^2\, dz^2\,.\label{metred}
\ee
Note that the two-dimensional field $\psi$ is redundant, in the sense
that it could be absorbed into a conformal rescaling of the two-dimensional
metric $ds_2^2$.  However, it it useful to retain it since it will allow
us later to take $ds_2^2$ to be just the Minkowski metric.  It is
straightforward to see that the three-dimensional Lagrangian (\ref{d3lag})
reduces to give
\be
{\cal L}_2 = \sqrt{-g}\, \rho\, \big(R-\ft14 g^{\mu\nu} \tr(A_\mu A_\nu) +
 \rho^{-1}\, g^{\mu\nu}  \del_\mu\rho\, \del_\nu\psi\big)\,,\label{d2lag}
\ee
(after an integration by parts).

   The equations of motion that follow from varying $\cV$, $\psi$\, $\rho$
and $g_{\mu\nu}$ are, respectively,
\bea
\nabla^\mu(\rho\, A_\mu) &=&0\,,\label{Veom}\\
\square\rho &=&0\,,\label{rhoeq}\\
\square\psi &=& R -\ft14 g^{\mu\nu}\, \tr(A_\mu A_\nu)\,,\label{psieq}\\\
0&=& \del_\mu\rho\, \del_\nu\psi + \del_\nu\rho\, \del_\mu\psi -g_{\mu\nu}
  \del_\sigma\rho\del^\sigma\psi - 2\nabla_\mu\nabla_\nu\rho\nn\\
&& -
  \ft12\tr \big(A_\mu A_\nu - 
   \ft12 g_{\mu\nu} \tr(A_\sigma A^\sigma)\big)\,,\label{einst}
\eea
(after using (\ref{rhoeq}) to simplify (\ref{einst}).

   Using general coordinate transformations, any two-dimensional metric
can be written, locally, as a conformal factor times the Minkowski metric.
Thus we may now take $ds_2^2$ to be the Minkowski metric, with $e^\psi$
as the required conformal factor.  It is convenient, furthermore, to
introduce light-cone coordinates $x^\pm$, so that we have 
$ds_2^2=2dx^+ dx^-$.  Note that the $++$ and $--$ components of the 
Einstein equation (\ref{einst}) can be used to solve for $\psi$, since
it gives
\be
(\del_+\rho)\, \del_+\psi= \del_+^2\rho + \ft14 \tr(A_+^2)\,,\qquad
(\del_-\rho)\, \del_-\psi= \del_-^2\rho + \ft14 \tr(A_-^2)\,.\label{psisol}
\ee
Equation (\ref{psieq}) can now be seen to be a consequence of (\ref{Veom}),
(\ref{rhoeq}), (\ref{psisol}) and (\ref{cartmau0}), so the two-dimensional
equations reduce to solving
\bea
d(\rho{*A}) &=&0\,,\label{eom}\\
\qquad dA+ A\wedge A&=&0\,,\label{cartmau}\\
\square\rho&=&0\,,\label{rhoeom}
\eea
with $\psi$ then being found using (\ref{psisol}).

\subsection{The Lax equation}\label{formsec}

   The use of the Lax equation in this context was discussed in \cite{pol},
and in a somewhat different, but related way, in \cite{nicolai}.
We shall begin by following Schwarz's discussion of the Lax-pair formulation 
in \cite{schwarz2}, except that we prefer
to use differential forms where possible, rather than light-cone
coordinates.  The equations (\ref{eom}) 
and (\ref{cartmau}) 
can both be derived from the integrability condition for a solution $X$ of
the Lax equation
\be
\tau (d+A) X= {*dX}\,.\label{Lax2}
\ee
By taking the appropriate linear
combination of this and its dual, we obtain\footnote{Note that $*^2=+1$ when
acting on 1-forms in signature $(1,1)$ spacetimes.  A summary of some further
useful properties of forms in two dimensions can be found in \cite{lupepo}.}
\be
dX X^{-1} = \fft{\tau}{1-\tau^2} \, {*A} + 
       \fft{\tau^2}{1-\tau^2} \, A\,.\label{Lax3}
\ee
Unlike the flat-space case where the spectral parameter $\tau$ is a constant,
here it must be allowed to have a specific dependence on the 
two-dimensional spacetime 
coordinates. It is convenient in what follows to parameterise $\tau$ in terms
of $\theta$ according to
\be
\tau=\tanh\ft12\theta\,,\label{tautheta}
\ee
which implies we shall also have
\be
s\equiv\sinh\theta = \fft{2\tau}{1-\tau^2}\,,\qquad
    c\equiv\cosh\theta = \fft{1+\tau^2}{1-\tau^2}\,.\label{scdef}
\ee

   In fact, one finds that the $x^\mu$ dependence of $\tau$ should
occur through the function $\rho(x^\mu)$, as follows:
\be
d\tau = \fft{\tau}{\rho}\, (c d\rho + s {*d}\rho)\,.\label{dtau}
\ee
 Equation (\ref{tautheta}) then implies
\be
d\theta = \fft{s}{\rho}\, (c d\rho + s {*d}\rho)\,.\label{dtheta}
\ee
With these preliminaries, it is now an elementary calculation to see
that from (\ref{Lax3}) that the Cartan-Maurer equation $d(dX X^{-1})=
  (dX X^{-1})\wedge (dX X^{-1})$ implies
\be
\tau\, (dA + A\wedge A) + \fft1{\rho}\, d(\rho {*A})=0\,,
\ee
from which the Bianchi identity (\ref{cartmau}) and the equation of
motion (\ref{eom}) indeed follow.

   A simple calculation also shows that the integrability condition 
that follows by taking the exterior derivative of (\ref{dtheta}) implies
\be
d{*d}\rho=0\,.\label{boxrho}
\ee
This is indeed the correct equation of motion (\ref{rhoeom}) 
for $\rho$.  If we now
introduce light-cone coordinates for the two-dimensional spacetime,
and work in the gauge where the metric is flat, $ds^2=2dx^+ dx^-$, then
on any function $f$ we shall have
\be
df=\del_+ f \, dx^+ + \del_-f\, dx^-\,,\qquad
{*d}f =\del_+ f \, dx^+ - \del_-f\, dx^-\,.
\ee
Thus (\ref{boxrho}) becomes $\del_+\del_-\rho=0$, with the general
solution
\be
\rho = \rho_+(x^+) + \rho_-(x^-)\,.\label{rhosol}
\ee

   The equation (\ref{dtheta}) that governs the $x^\mu$ dependence of
$\theta$ becomes, using the light-cone coordinates,
\be
\del_+\theta = \fft{1}{2\rho}\, (e^{2\theta} -1)\, \del_+\rho\,,\qquad
  \del_-\theta = \fft{1}{2\rho}\, (1-e^{-2\theta})\, \del_-\rho\,.
\ee
These equations may be integrated to give
\be
1-e^{-2\theta} = \rho\, f_-(x^-)\,,\qquad 
e^{2\theta} -1 = \rho\, f_+(x^+)\,,\label{thetasol}
\ee
where $f_\pm$ are arbitrary functions of their respective arguments.  
Eliminating $\theta$ between the two expressions implies
\be
\rho= \fft1{f_-(x^-)} - \fft1{f_+(x^+)}\,.
\ee
Comparing with (\ref{rhosol}), we can write
\be
\fft1{f_-(x^-)} = \rho_-(x^-) + \fft1{2\lambda} \,,\qquad
\fft1{f_+(x^+)} = -\rho_+(x^+) + \fft1{2\lambda}\,,
\ee
where $\lambda$ is a constant.  From (\ref{rhosol}) and (\ref{thetasol}),
the solution for $\theta$ is given by
\be
e^{2\theta} = \fft{1+2\lambda\, \rho_-(x^-)}{1-2\lambda\, \rho_+(x^+)}\,.
\label{thetasol2}
\ee
It then follows from (\ref{tautheta}) that
\bea
\tau &=& \fft1{\lambda\rho}\, \Big[ 1- \lambda(\rho_+ -\rho_-) -
  \sqrt{(1+2\lambda\rho_-)(1-2\lambda\rho_+)}\Big] \,,\label{taulambda}\\
&=& \ft12\lambda \rho + \ft12\lambda^2\rho(\rho_+-\rho_-) +
   \ft18\lambda^3 \rho[\rho^2 + 4(\rho_+-\rho_-)^2] +\cdots\,.\nn
\eea
Note that we have chosen the negative sign in front of the square root,
so that $\tau$ is analytic in $\lambda$ in the neighbourhood of
$\lambda=0$.  Furthermore, $\tau=0$ when $\lambda=0$.

 In what follows, the constant parameter $\lambda$ will play the r\^ole of the
spectral parameter for the Lax equation.  The solution $X$ of the Lax 
equation (\ref{Lax2}) depends on the spectral function $\tau$, and since 
$\tau$ is given by (\ref{taulambda}) it follows that we may characterise
$X$ by the value of the arbitrary constant spectral parameter $\lambda$.
Thus when we consider a solution $X$ of the Lax equation, we shall 
denote it by $X(\lambda)$.  
(Of course, $X$ also depends on the two-dimensional
spacetime coordinates $x^\mu$, but we shall suppress the explicit 
indication of this dependence.) 
We shall choose the solution so that $X(0)=1$;
this is clearly consistent with the Lax equation (\ref{Lax3}), since
$\tau$ vanishes at $\lambda=0$.

\section{The Kac-Moody Symmetries}\label{kac-moodysec}

\subsection{The Kac-Moody transformations}

   The symmetries of the two-dimensional system are described by 
field transformations that preserve the equations of motion (\ref{eom})
and (\ref{rhoeom}).  The full set of symmetries will involve transformations
of the coset representative $\cV$ that are expressed using the 
solution $X$ of the Lax equation (\ref{Lax2}).  It is therefore necessary
also to find how $X$ itself transforms; this is determined by requiring in
addition that the Lax equation (\ref{Lax2}) be left invariant.

  We begin, following \cite{schwarz1,schwarz2}, by constructing transformations
of $\cV$.  These are of the form 
\be
\delta \cV = w \cV \eta + \delta h \cV\,,\qquad \eta\equiv 
   X(\lambda)\ep X(\lambda)^{-1}\,,\label{delcV0}
\ee
where $\ep$ an infinitesimal constant parameter taking values in the
Lie algebra $\cG$, and $\delta h$ is a $\lambda$-dependent and 
field-dependent compensating transformation in $\cH$ that restores the original
gauge.  The quantity $w$ is a function that is a singlet under
the Lie-algebra, and will be determined shortly. 
The meaning of (\ref{delcV0}) is as follows.  The
coset representative $\cV$ itself is, of course, independent of
the spectral parameter $\lambda$.  The transformation $\delta$ is
$\lambda$-dependent, and in fact we expand it as a power series
in $\lambda$:
\be
\delta(\ep,\lambda) = \sum_{n\ge0} \lambda^n\, \delta_\n(\ep)\,.
\ee
By equating coefficients of each power of $\lambda$ in the Taylor expansions
of the two sides of equation (\ref{delcV0}), one therefore obtains a 
hierarchy of transformations $\delta_\n(\ep)\cV$.  
One can take independent $\ep$ parameters
for each $n$.  The transformations at the $n=0$ order are just the original
infinitesimal global $\cG$ symmetries.

   In the flat-space case discussed in 
\cite{schwarz1,lupepo} $w$ is spacetime-independent, and one can take $w=1$, 
but in the gravity-coupled situation we are considering in this paper,
one must choose a very specific $x^\mu$-dependence for $w$ in order
to ensure that (\ref{delcV0}) describes a symmetry of the equation
of motion $d(\rho{*A})=0$.  The calculation is performed by first noting
from (\ref{MAdef}) that (\ref{delcV0}) implies
\be
\delta M = w M\eta + w \eta^\dagger M\,,\qquad 
\delta A= D(w\eta) + D(w M^{-1} \eta^\dagger M)\,,\label{delMA}
\ee
where we define $Df\equiv df + [A,f]$ on any $\cG$-valued function $f$.
Note that $\rho$ will be inert under the Kac-Moody transformations.

    Use of the Lax equation (\ref{Lax2}) shows that
\be
D\eta= \fft1{\tau}\, {*d\eta}\,,\qquad
 D(M^{-1}\eta^\dagger M)= \tau\, {*d}(M^{-1}\eta^\dagger M)\,,
\ee
as in the flat-space case \cite{lupepo}. From this, it follows that
\bea
d(\rho {*\delta A}) &=& \Big[d\Big(\fft{\rho w}{\tau}\Big)-
     \rho {*d}w \Big]\wedge d\eta + 
 \big[d(\rho w \tau)- \rho {*d}w\big]\wedge d(M^{-1}\eta^\dagger M)\nn\\
&& +
d(\rho{*d}w)\, \big(\eta + M^{-1}\eta^\dagger M\big)\,.
\eea
and so $d(\rho {*\delta A})$ will vanish (and thus the equation of
motion (\ref{eom}) will be preserved by the transformations) if 
\cite{schwarz2}
\be
\rho {*dw} = d\Big(\fft{\rho w}{\tau}\Big)\,,\qquad 
   \hbox{and}\qquad \rho{*dw} = d(\rho w \tau)\,.
\ee
Using (\ref{dtheta}) one sees that this can be achieved, by taking $w$
to be a constant multiple of $s/\rho$.  
We shall take $w=s/(\lambda\rho)$, and thus the 
transformations (\ref{delcV0}) we shall consider are given by
\be
\delta \cV = \fft{s}{\lambda\rho}\, \cV\eta + \delta h \cV\,,\qquad
\eta\equiv X(\lambda)\ep X(\lambda)^{-1}\,.\label{delcV}
\ee
(The reason for choosing to divide out by the constant parameter $\lambda$
is that then, as can be seen from (\ref{scdef}) and (\ref{taulambda}), 
the prefactor $w=s/(\lambda\rho)$ approaches 1 as $\lambda$ goes to zero.)

   We now need to consider the transformations of $X$.  These are determined
by the requirement that the Lax equation must be preserved, with $A$
transforming as in (\ref{delMA}).  We shall need to know how $X(\lambda)$ with
spectral parameter $\lambda_2$ transforms under variations 
$\delta(\ep,\lambda)$ with an independent choice of spectral parameter
$\lambda_1$.  We denote $X(\lambda_2)$ by $X_2$, and $\delta(\ep_1,\lambda_1)$
by $\delta_1$.  From (\ref{Lax3}) we therefore have
\be
\delta_1(dX_2 X_2^{-1}) = \fft{\tau_2}{1-\tau_2^2}\, {*\delta_1 A} + 
         \fft{\tau_2^2}{1-\tau_2^2}\, \delta_1 A\,.\label{varlax}
\ee
($\tau_2$ means the solution for $\tau$ given in (\ref{taulambda}) with
the spectral parameter $\lambda$ taken to be $\lambda_2$.) 

   Let us write the transformation of $X$ as $\delta_1 X_2 = U X_2$.  It 
follows that $\delta_1(dX_2 X_2^{-1})=dU + [U, dX_2 X_2^{-1}]$, and so 
from (\ref{varlax}) we have
\be
dU + [U,dX_2 X_2^{-1}] = \fft{\tau_2}{1-\tau_2^2}\, {*\delta_1 A} +
         \fft{\tau_2^2}{1-\tau_2^2}\, \delta_1 A\,.\label{varlax2}
\ee
We can write the solution for $U$ as the sum 
\be
  U= U^{\rm{hom}} + U^{\rm{inhom}}\,,
\ee
where $U^{\rm{hom}}$ is the solution of the homogeneous equation with
the right-hand side of (\ref{varlax2}) set to zero, while $ U^{\rm{inhom}}$
solves the inhomogeneous equation with the variations of $A$ on the
right-hand side providing the source.

   It is easily seen that the homogeneous equation is solved by
\be
U^{\rm{hom}} = X_2\ep_1 X_2^{-1}\,.\label{homsol}
\ee
For the inhomogeneous equation, we make the ansatz
\be
U^{\rm{inhom}} = u\, \eta_1  + v\, M^{-1}\eta_1^\dagger M \,,
\ee
where $u$ and $v$ are $\cG$-singlet functions to be determined.  Note that
$\eta_1\equiv X_1 \ep_1 X_1^{-1}$, with $X_1\equiv X(\lambda_1)$, 
and that therefore
$d\eta_1 = -[\eta_1, dX_1 X_1^{-1}]$.  Substituting the variations of
$A$, and the ansatz for $U=U^{\rm{inhom}}$, into (\ref{varlax2}) one
finds, after some straightforward although slightly elaborate algebra
involving the use of the Lax equation (\ref{Lax3}), (\ref{dtau}) and 
(\ref{dtheta}), that there is a solution with \cite{schwarz2}
\be
u= \fft{s_1}{\lambda_1 \rho}\, \fft{\tau_2}{(\tau_1-\tau_2)}\,,\qquad
 v= \fft{s_1}{\lambda_1\rho}\, \fft{\tau_1\tau_2}{(1-\tau_1\tau_2)}\,.
\label{uvsol}
\ee
Here $s_1$ denotes $s=\sinh\theta=2\tau/(1-\tau^2)$ with $\tau$ given by
(\ref{taulambda}) for $\lambda=\lambda_1$.
A remarkable property, which is absolutely crucial in what follows, is
that $d(u-v)=0$ and hence
\be
u-v=\hbox{constant}\,.\label{uvconst}
\ee
This can be verified using (\ref{dtau}) and (\ref{dtheta}).

   To summarise the results so far, we have shown that there exist symmetry
transformations of $X$ of the form
\bea
\delta_1^{\rm{hom}} X_2 &=& X_2 \ep_1\,,\label{deltahom}\\
\delta_1^{\rm{inhom}} X_2 &=& \fft{s_1}{\lambda_1 \rho}\, 
   \fft{\tau_2}{(\tau_1-\tau_2)}\, \eta_1 X_2 + 
    \fft{s_1}{\lambda_1\rho}\, \fft{\tau_1\tau_2}{(1-\tau_1\tau_2)}\,
   M^{-1}\eta_1^\dagger M X_2\,.\label{deltainhom}
\eea
The inhomogeneous transformations are accompanied by the transformations
of $\cV$, $M$ and $A$ given by (\ref{delcV}) and (\ref{delMA}), 
whilst the homogeneous transformations
are completely independent, acting only on $X$ and leaving $\cV$, $M$
and $A$ invariant. 

   The complete set of global symmetry transformations of $X$ will be read off
by expanding the various $\delta_1$ variations, and $X_2$, as power series
in $\lambda_1$ and $\lambda_2$ respectively, both around $\lambda_i=0$.
Since $\tau_i$ vanishes at $\lambda_i=0$ (see (\ref{taulambda})) 
this appears to present a difficulty for
the inhomogeneous transformations (\ref{deltainhom}), because of the pole
associated with the 
denominator $(\tau_1-\tau_2)$ in the first term.  In the analogous
treatment of the flat-space case (see \cite{schwarz1,lupepo}) this was
not a problem, because there $\tau_1$ and $\tau_2$ were themselves constant 
spectral parameters and so the pole could subtracted by including an
appropriate constant multiple of the homogeneous transformation.  Here,
we cannot simply subtract the analogous multiple of $\delta^{\rm{hom}}$ 
with a prefactor of the form $(s_1/\lambda_1\rho)\, 
  \tau_2(\tau_1-\tau_2)^{-1}$, 
because this is not a constant, and so $(s_1/\lambda_1\rho)\, 
\tau_2(\tau_1-\tau_2)^{-1} 
\delta_1^{\rm{hom}}X_2$ is not a symmetry transformation.  At this
point the remarkable property (\ref{uvconst}) comes to the rescue.  In fact,
by substituting (\ref{tautheta}) and (\ref{thetasol2}) into (\ref{uvsol})
we find that
\be
\fft{s_1}{\lambda_1 \rho}\, \fft{\tau_2}{(\tau_1-\tau_2)} =
\fft{s_1}{\lambda_1\rho}\, \fft{\tau_1\tau_2}{(1-\tau_1\tau_2)}
  + \fft{\lambda_2}{\lambda_1-\lambda_2}\,.\label{uvrel2}
\ee
Thus, we may rewrite (\ref{deltainhom}) as
\be
\delta_1^{\rm{inhom}} X_2 = \fft{s_1}{\lambda_1\rho}\, 
  \fft{\tau_1\tau_2}{(1-\tau_1\tau_2)}\, \big(
 \eta_1 + M^{-1}\eta_1^\dagger M\big) X_2 + 
  \fft{\lambda_2}{\lambda_1-\lambda_2}\, \eta_1 X_2\,.\label{deltainhom2}
\ee
This shows that, despite the original appearance in (\ref{deltainhom}), 
the pole at 
$\lambda_1=\lambda_2$ in the inhomogeneous transformation rule actually
has a pure constant coefficient, and so by
subtracting the homogeneous symmetry transformation 
$\lambda_2/(\lambda_1-\lambda_2)\, \delta_1^{\rm{hom}}$ we can easily
remove the pole.\footnote{In Schwarz's discussion in \cite{schwarz2}, the 
important point that one must remove the $\tau_1=\tau_2$ singularity in 
$\delta_1^{\rm{inhom}}X_2$, and furthermore that this can actually be done,
for all spacetime points $x^\mu$ simultaneously, appears to have been 
unnoticed.  In fact Schwarz instead made a subtraction such that 
$\delta_1^{\rm{inhom}}X_2$ vanished at a preferred point $x_0^\mu$ in 
the two-dimensional spacetime.  However, this subtraction does not in fact
cancel the singularity at $\tau_1=\tau_2$ for other points in the 
spacetime.  The absence of the cancellation did not show up in Schwarz's
subsequent calculation of the commutator $[\delta_1,\delta_2]$ because
he evaluated it only on $\cV$ (which is inert under the homogeneous
variation in question)
and not on $X_3$ (which is not inert).}

   With these points understood, we can now present the full set of global
symmetry transformations of the two-dimensional system in the final
form that we shall use in what follows.  We shall denote them by
$\delta_1$ and $\td\delta_1$, and their actions on the original
sigma model fields in $\cV$ (and hence on $M$), 
and on the fields in $X(\lambda)$, are as follows:
\bea
\delta_1 \cV &=& \fft{s_1}{\lambda_1\rho}\, \cV\eta_1 + \delta h \cV\,,
\label{deltacV}\\
\delta_1 M &=&  \fft{s_1}{\lambda_1\rho}\, 
  \big(M\eta_1 + \eta_1^\dagger M\big)\,,\label{deltaM}\\
\delta_1 X_2 &=& \fft{\lambda_2}{\lambda_1-\lambda_2}\, 
  \big(\eta_1 X_2 - X_2 \ep_1\big) + \fft{s_1}{\lambda_1\rho}\,
   \fft{\tau_1\tau_2}{1-\tau_1\tau_2}\, 
            \big(\eta_1 + M^{-1}\eta_1^\dagger M\big) X_2\,,\label{deltaX}
\eea
where $\eta_1\equiv X_1\ep_1 X_1^{-1}$, and 
\bea
\td\delta_1 \cV &=&0\,,\label{tddeltacV}\\
\td\delta_1 M &=&0\,,\label{tddeltaM}\\
\td\delta_1 X_2 &=& \fft{\lambda_1\lambda_2}{1-\lambda_1\lambda_2}\, 
       X_2\ep_1\,.\label{tddeltaX}
\eea
The $\delta$ transformations are the inhomogeneous transformations we
discussed above, with the subtraction of the necessary homogeneous term 
in $\delta_1 X_2$ to ensure analyticity when $\lambda_1$ approaches 
$\lambda_2$.  The $\td\delta$ transformations are 
independent and purely homogeneous, thus leaving $\cV$ and $M$ invariant.  The
inclusion of the specific $\lambda_i$-dependent prefactor in
(\ref{tddeltaX}) is purely for convenience; it ensures that the final
algebra obtained by calculating the commutators of the transformations
arises in a simple and conventional basis.  

   It is evident from the
transformations above that the expansions of the variations $\delta$ and
$\td\delta$ will be of the forms
\be
\delta(\ep,\lambda) = \sum_{n\ge0} \lambda^n\, \delta_\n(\ep)\,,\qquad
\td\delta(\ep,\lambda) = \sum_{n\ge1} \lambda^n\, \td\delta_\n(\ep)\,.
\label{deltaexps}
\ee
There is an independent $\cG$-valued infinitesimal parameter for each 
$n$ in $\delta_\n(\ep)$, and for each $n$ in $\td\delta_\n(\ep)$.

\subsection{The Kac-Moody algebra}

   Having obtained the explicit expressions (\ref{deltacV})--(\ref{tddeltaX})
for the $\delta$ and $\td\delta$ transformations of the fields, it is
now a mechanical exercise to calculate the commutators of these 
transformations.  Specifically, we calculate the commutators 
$[\delta_1,\delta_2]$, $[\delta_1,\td\delta_2]$ and 
$[\td\delta_1,\td\delta_2]$ acting on $M$ and on $X_3$.  After some algebra,
we find
\bea
{[}\delta_1,\delta_2{]} &=&
 \fft{\lambda_1}{\lambda_1-\lambda_2}\,
      \delta(\ep_{12}, \lambda_1)- \fft{\lambda_2}{\lambda_1-\lambda_2}\,
   \delta(\ep_{12},\lambda_2)\,,
\label{nn}\\
{[}\delta_1,\td\delta_2{]} &=& 
  \fft{\lambda_1 \lambda_2}{1-\lambda_1 \lambda_2}\,
               \delta(\ep_{12},\lambda_1) + \fft1{1-\lambda_1 \lambda_2}\,
          \td\delta(\ep_{12},\lambda_2)\,,\label{ntd}\\
{[}\td\delta_1,\td\delta_2{]} &=& \fft{\lambda_2}{\lambda_1-\lambda_2}\,
  \td\delta(\ep_{12},\lambda_1)
  -\fft{\lambda_1}{\lambda_1-\lambda_2}\, 
         \td\delta(\ep_{12}, \lambda_2)\,,\label{tdtd}
\eea
where $\ep_{12}\equiv [\ep_1,\ep_2]$.  Note that there are no poles
at $\lambda_1=\lambda_2$, because in (\ref{nn}) and (\ref{tdtd}) the numerator
on the right-hand side has a zero that cancels the denominator there.

   Using (\ref{deltaexps}), expanding the expressions 
(\ref{nn})--(\ref{tdtd}) in powers of $\lambda_1$ and $\lambda_2$, and
then equating the coefficients of each power, we can read off the algebra
of the modes, finding
\bea
{[}\delta_\m(\ep_1),\delta_\n(\ep_2){]} &=& \delta_{\sst{(m+n)}}(\ep_{12})\,,
\qquad m\ge0\,,\ n\ge0\,,\label{nn2}\\
{[}\delta_\m(\ep_1),\td\delta_\n(\ep_2){]} &=& \delta_{\sst{(m-n)}}(\ep_{12})
        + \td\delta_{\sst{(n-m)}}(\ep_{12})\,,\qquad m\ge0\,,\ n\ge1\,,
   \label{ntd2}\\
 {[}\td\delta_\m(\ep_1),\td\delta_\n(\ep_2){]} &=&
       \td\delta_{\sst{(m+n)}}(\ep_{12})\,,
  \qquad m\ge1\,,\ n\ge1\,,\label{tdtd2}
\eea
where in (\ref{ntd2}) it is to be understood that $\delta_\n=0$ for
$n\le -1$ and $\td\delta_\n=0$ for $n\le0$.

   As in the flat-space case discussed in \cite{lupepo}, these three sets
of commutation relations can be combined into one
by introducing a new set $\Delta_\n$ of variations, defined for all $n$
with $-\infty\le n\le \infty$, according to
\bea
   \Delta_\n &=& \delta_\n\,,\qquad n\ge0\,,\nn\\
 \Delta_{\sst{(-n)}} &=& \td\delta_\n\,,\qquad n\ge 1\,.\label{Deltadef}
\eea
It is then easily seen that (\ref{nn2}), (\ref{ntd2}) and (\ref{tdtd2})
become
\be
[\Delta_\m(\ep_1),\Delta_\n(\ep_2)] = \Delta_{\sst{(m+n)}}(\ep_{12})\,,
\qquad m,n\in\Z \,,\label{kacmoody}
\ee
with $\ep_{12}=[\ep_1,\ep_2]$.
This defines the affine Kac-Moody algebra $\hat \cG$.  If we associate
generators $J_n^i$ with the transformations $\Delta_\n(\ep^i)$, where
$\ep= \ep^i\, T^i$ and $T^i$ are the generators of the Lie algebra $\cG$
satisfying $[T^i,T^j]=f^{ij}{}_k\, T^k$, then (\ref{kacmoody}) implies
\be
[J_m^i,J_n^j]= f^{ij}{}_k\, J^k_{m+n}\,.\label{kacmoody2}
\ee

\section{Virasoro-Like Symmetries}\label{virasorosec}

\subsection{Virasoro-like transformations}\label{virsubsec}

     In addition to the Kac-Moody symmetries that we discussed in section
\ref{kac-moodysec}, which are the affine extension of the original
$\cG$ global symmetry of the sigma model, there is also a further
infinite-dimensional symmetry that is a singlet under $\cG$ 
\cite{houli,maison,nicolai}.   This is related to the Virasoro algebra. 
It was discussed in detail for sigma models in flat two-dimensional
spacetime in \cite{schwarz1}, but the generalisation to the gravity-coupled
case was not found in \cite{schwarz2}.  Here, we use the methods developed
in \cite{schwarz1,schwarz2}, and extend them to obtain explicit Virasoro-like
transformations in the gravity-coupled sigma models.

   The action of the Virasoro-like transformations on the coset representative
$\cV$ will be taken to be\footnote{We should really include an infinitesimal 
parameter as a prefactor in the definition of $\xi$ in
equation (\ref{deltaVcV}).  However, since it is a singlet
it plays no significant r\^ole, and so it may be omitted without any
risk of ambiguity.} 
\be
\delta^V \cV = h\cV \xi\,,\qquad \xi \equiv \dot X X^{-1}\,,
\label{deltaVcV}
\ee
where $h$ is a $\cG$-singlet function to be determined, and $\dot X$ means
$dX(\lambda)/d\lambda$.  From this it follows that
\be
\delta^V A = D(h(\xi + M^{-1} \xi^\dagger M))\,.\label{deltaVA}
\ee
We can show from (\ref{thetasol2}) that
\be
\dot\theta = \fft{s^2}{\lambda^2 \rho}\,,\qquad
\dot\tau= \fft{s\tau}{\lambda^2\rho}\,.
\ee
By taking the $\lambda$ derivative of the Lax equation (\ref{Lax3}),
we can then show after a little algebra that
\be
D\xi= \fft1{\tau}\, {*d}\xi -\fft{s^2}{2\lambda^2\rho\tau}\, (\tau {*A} +A)\,,
\qquad 
D(M^{-1}\xi^\dagger M)= \tau {*d}(M^{-1} \xi^\dagger M) +
  \fft{s^2}{2\lambda^2\rho}\, ({*A}+\tau A)\,,
\ee
and hence that
\be
D(\xi + M^{-1}\xi^\dagger M)=\fft1{\tau}\, {*d}\xi + 
  \tau {*d}(M^{-1} \xi^\dagger M) -\fft{s}{\lambda^2\rho}\, A\,.\label{virid}
\ee

   We now examine the variation of the equation of motion for $A$,
namely 
\be
\delta^V\big(d(\rho {*A})\big)= 
d\big( (\delta^V\rho){*A} + \rho {*\delta^V A}\big) =0\,.\label{eomvar}
\ee
Note that unlike the Kac-Moody transformations, which leave $\rho$ invariant,
here we shall find that the Virasoro-like transformations must act also
on $\rho$.   Substituting (\ref{deltaVA}) into (\ref{eomvar}), and
using (\ref{virid}), we obtain
\bea
\delta^V\big(d(\rho {*A})\big) &=&
\Big[d\Big(\fft{\rho h}{\tau}\Big) - \rho {*dh}\Big]\wedge d\xi +
 [d(\rho h\tau) - \rho{*dh}]\wedge d(M^{-1}\xi^\dagger M)\nn\\
&& +d(\rho {*dh})[\xi + M^{-1}\xi^\dagger M] -d\Big(\fft{h s}{\lambda^2}\,
  {*A}\Big) + d(\delta^V\rho\, {*A})\,.\label{var2}
\eea
This leads us to require $h$ to satisfy
\be
d\Big(\fft{\rho h}{\tau}\Big) =d(\rho h\tau)=\rho{*dh}\,,
\ee
which has as solution
\be
   h=-\fft{s}{\rho}\,,\label{hsol2}
\ee
(the constant factor is arbitrary, and we take it to be $-1$ for
later convenience).  Equation 
(\ref{var2}) now reduces to 
\be
\delta^V\big(d(\rho {*A})\big)= d\Big(\fft{s^2}{\lambda^2\rho}\,
  {*A}\Big) + d(\delta^V\rho\, {*A})  \,,
\ee
which, in view of the fact that $d(\rho {*A})=0$, vanishes if we take
\be
\delta^V\rho = -\fft{s^2}{\lambda^2\rho} +\alpha \rho\,,\label{deltarho}
\ee
where $\alpha$ is an arbitrary constant parameter.  However, since
by its definition $\delta^V$ acting on $\cV$ has no $\lambda^0$ term
in its Taylor expansion, we may choose $\alpha$ so as to remove the
$\lambda^0$ term in (\ref{deltarho}).  (We shall return to discussing 
the extra symmetry associated with $\alpha$ later.)  This means setting
\be
\alpha=1\,.
\ee

   Since $\rho$ satisfies the free wave equation (\ref{rhoeq}), we must
also check that its variation (\ref{deltarho}) is consistent with this.
This is easily done, by substituting (\ref{thetasol2}) into (\ref{deltarho}),
giving
\be
\delta^V \rho = -\fft{2\lambda\rho_+^2}{1-2\lambda \rho_+} +
   \fft{2\lambda \rho_-^2}{1+2\lambda \rho_-} \,.\label{deltarho2}
\ee
Since this is manifestly the sum of a function of $\rho_+(x^+)$ and a
function of $\rho_-(x^-)$, it clearly satisfies 
$\del_+\del_-(\delta^V\rho)=0$.  In fact we can read off from 
(\ref{deltarho2}) that
\be
\delta^V\rho_+ = -\fft{2\lambda\rho_+^2}{1-2\lambda\rho_+} + \beta\,,\qquad
\delta^V\rho_- = \fft{2\lambda\rho_-^2}{1+2\lambda\rho_-} -\beta\,,
\label{deltarhoprhom}
\ee
where $\beta$ is another arbitrary constant parameter.  Again, like
the previous discussion which implied we could take $\alpha=1$, here
 too we could require that there be no $\lambda^0$ term in the expansion, 
and this constrains $\beta$ also, to 
\be
\beta=0\,.
\ee
We shall discuss the extra symmetry associated with a non-zero $\beta$
below.

   It is useful also to record the expressions for $\delta_1^V\theta_2$ and
$\delta_1^V\tau_2$, for which we find
\bea
\delta_1^V \theta_2 &=& -\fft{2 \tau_1\tau_2}{\rho^2 
      (\t_1-\tau_2)(1-\tau_1 \tau_2)}\, \Big(\fft{s_1^2}{\lambda_1^2}
               -\fft{s_2^2}{\lambda_2^2}\Big)\,,\label{delVtheta}\\
\delta_1^V \tau_2 &=& - \fft{\tau_1\tau_2(1-\tau_2^2)}{\rho^2
      (\t_1-\tau_2)(1-\tau_1 \tau_2)}\, \Big(\fft{s_1^2}{\lambda_1^2}
               -\fft{s_2^2}{\lambda_2^2}\Big)\,,\label{delVtau}
\eea

   The next step is to calculate $\delta_1^V X_2$.  This is done by
requiring the invariance of the Lax equation (\ref{Lax3}) 
(for $\lambda=\lambda_2$) under the transformation $\delta_1^V$,
with $\delta_1^V A$ and $\delta_1^V\tau_2$ read off from (\ref{deltaVA})
and (\ref{delVtau}).  After some algebra, we find
\be
\delta_1^V X_2 = -\fft{s_1}{\rho} \fft{\tau_1\tau_2}{(1-\tau_1\tau_2)}\,
(\xi_1 + M^{-1} \xi_1^\dagger M) X_2 - 
  \fft{\lambda_1\lambda_2}{\lambda_1-\lambda_2}\, (\xi_1-\xi_2) X_2\,.
\label{deltaVX}
\ee

\subsection{Virasoro-like symmetries}

   We are now in a position to calculate the commutator
$[\delta_1^V,\delta_2^V]$.  We have evaluated this on all the
fields, namely $\cV$, $\rho$ and $X$.   We find that the abstract algebra
turns out to be
\be
[\delta_1^V,\delta_2^V] = 
\fft{\lambda_1\lambda_2}{\lambda_1-\lambda_2}\,
(\dot\delta_1^V + \dot\delta_2^V)
-\fft{2\lambda_1\lambda_2}{(\lambda_1-\lambda_2)^2}\,
(\delta_1^V -\delta_2^V)
\,.\label{valg1}
\ee

  Making the mode expansion
\be
\delta^V(\lambda) = \sum_{n\ge1} \lambda^n \delta^V_\n\,,
\label{virmodes}
\ee
we find upon substituting into (\ref{valg1}) that
the modes satisfy the commutation relations
\be
[\delta^V_\m,\delta^V_\n] = (m-n) \delta^V_{\sst{(m+n)}}\,,\qquad m\ge1\,\ \ 
   n\ge 1\,.\label{virexp1}
\ee
As in the flat-space case we have obtained ``half'' the Virasoro algebra.
It is interesting that here we obtain precisely the upper half of a standard
Virasoro algebra of the $L_n$, whereas in the flat-space case we found 
generators $K_n=L_n-L_{-n}$.  The difference between the two results from
the use of a different spectral parameterisation.  (See appendix \ref{appsec} 
for a
discussion of how the flat-space limit of our gravity-coupled construction
here is related to the previous flat-space construction in \cite{lupepo}.)

   We now return to the additional symmetries associated with the 
parameter $\alpha$ in (\ref{deltarho}) and $\beta$ in (\ref{deltarhoprhom}). 
It will be recalled that we ``fixed'' these symmetries by requiring the
absence of $\lambda^0$ terms in $\delta^V\rho$ and in $\delta^V\rho_\pm$,
on the grounds that the original Virasoro-like transformations $\delta^V\cV$
had contributions only for strictly positive powers of $\lambda$.  However,
we can interpret the transformations parameterised by $\alpha$
and $\beta$ as additional symmetries of the system that act non-trivially
on $\rho$, but which happens to leave $\cV$ inert.  In fact it is
natural to denote the $\alpha$ symmetry by $\hat\delta^V_{\sst{(0)}}$ and
the $\beta$ symmetry by $\hat\delta^V_{\sst{(-1)}}$, since we find that they
are the natural continuation of the Virasoro transformations $\delta^V_\n$
in (\ref{virexp1}) to include the $n=0$ and $n=-1$ terms.  

   In detail, we find that the action of the $\hat\delta^V_{\sst{(0)}}$ and
$\hat\delta^V_{\sst{(-1)}}$ transformations are as follows:
\bea
\hat\delta^V_{\sst{(0)}}\rho_+ &=&  -\rho_+\,,\qquad 
      \hat\delta^V_{\sst{(0)}} \rho_- = -\rho_-\,,\qquad 
\hat\delta^V_{\sst{(0)}} X = -\lambda X(\lambda)\,, 
\qquad \hat\delta^V_{\sst{(0)}} \cV =0\,, \nn\\
\hat\delta^V_{\sst{(-1)}}\rho_+ &=&  -\ft12 \,,\qquad
      \delta^V_{\sst{(-1)}} \rho_- = \ft12 \,,\qquad
\hat\delta^V_{\sst{(-1)}} X = -\lambda^2 X(\lambda)\,,
\qquad \hat\delta^V_{\sst{(-1)}} \cV =0\,,\label{virextra}
\eea
It is also useful to record that
\be
\hat\delta^V_{\sst{(0)}}\rho= -\rho\,,\quad \hat\delta^V_{\sst{(0)}}\tau=
   -\lambda \dot\tau\,;\qquad
\hat\delta^V_{\sst{(-1)}}\rho= 0\,,\quad \hat\delta^V_{\sst{(-1)}}\tau=
   -\lambda^2 \dot\tau\,.
\ee

   It is a straightforward matter to calculate the commutators of 
$\hat\delta^V_{\sst{(0)}}$ and $\hat\delta^V_{\sst{(-1)}}$ with each other, and
with the previously-defined Virasoro-type transformations appearing in
(\ref{virmodes}).  We first define a total extended set of transformations 
$\hat\delta^V_\n$ with $n\ge -1$, according to
\be
\hat\delta^V_\n = \delta^V_\n\,,\qquad n\ge 1\,,
\ee
where $\delta^V_\n$ is defined in (\ref{virmodes}), and with 
$\hat\delta^V_\n$ for $n=0$ and $n=-1$ as in (\ref{virextra}).  We find
that these satisfy precisely the same algebra as in (\ref{virexp1}),
now with the index range extended appropriately, {\it viz.}
\be
[\hat\delta^V_\m,\hat\delta^V_\n] = (m-n) 
   \hat\delta^V_{\sst{(m+n)}}\,,\qquad m\ge-1\,\ \
   n\ge -1\,.\label{virexp2}
\ee
Thus we may associate standard Virasoro generators $L_n$ with 
$\hat\delta^V_\n$,
and we obtain the Virasoro subalgebra generated by $L_n$ with $n\ge -1$:
\be
[L_m,L_n]=(m-n) L_{m+n}\,,\qquad m\ge 1\,,\ n\ge1\,.
\ee

   It is natural to introduce an extended Virasoro-like transformation
$\hat\delta^V$ constructed from the extended mode set 
$\hat\delta^V_\n$, by analogy
with (\ref{virmodes}).  Thus we may define
\be
\hat\delta^V(\lambda) = \sum_{n\ge -1} \lambda^n \hat\delta^V_\n =
   \delta^V(\lambda) + \hat\delta^V_{\sst{(0)}} + \fft{1}{\lambda}\,
      \hat\delta^V_{\sst{(-1)}}\,.
\ee
Acting, for example, on $\rho_+$, we find from (\ref{deltarhoprhom}) 
(with $\beta=0$) and (\ref{virextra}) that
\be
\hat\delta^V \rho_+= -\fft{2\lambda\rho_+}{1-2\lambda\rho_+} -
  \rho_+ - \fft1{2\lambda} = -\fft{1}{2\lambda(1-2\lambda\rho_+)}\,.
\ee
The $\hat\delta^V$ variations of $\rho_-$, $\rho$ and $X$ can all easily
be worked out in a similar way from our previous results.  ($\cV$, as
we noted already, is inert under the extra terms $ \hat\delta^V_{\sst{(0)}}$
and $\hat\delta^V_{\sst{(-1)}}$.)  In summary, the extended transformations
$\hat\delta^V$ act on all the fields as follows:
\bea
\hat\delta^V \rho_+ &=& -\fft1{2\lambda(1-2\lambda\rho_+)}\,,\qquad
 \hat\delta^V \rho_- = \fft1{2\lambda(1+2\lambda\rho_-)}\,,\qquad
\hat\delta^V \rho = -\fft{s^2}{\lambda^2\rho}\,,\nn\\
\hat\delta_1^V X_2&=& -\fft{s_1}{\rho}\, \fft{\tau_1\tau_2}{(1-\tau_1\tau_2)}\,
 \big(\xi_1 + M^{-1}\xi_1^\dagger M\big) X_2 - 
\fft{\lambda_2 /\lambda_1}{(\lambda_1-\lambda_2)}\, 
  \big(\lambda_1^2\, \xi_1 -\lambda_2^2 \, \xi_2\big)\,,\nn\\
\hat\delta^V \cV &=& -\fft{s}{\rho}\, \cV \xi\,.\label{finalV}
\eea

\subsection{Commutators of Virasoro and Kac-Moody transformations}

   Having obtained the Kac-Moody transformations in section \ref{kac-moodysec},
and the ``half'' Virasoro transformations in section \ref{virsubsec}, we may
now compute also the mixed commutators between these two sets of symmetry
transformations.  The calculations are entirely mechanical, although in
some cases somewhat involved, and we shall just present our final results. 

   For the Virasoro-like transformations $\delta^V$ defined in 
(\ref{deltaVcV}), (\ref{deltarho}) and (\ref{deltaVX}), we find that their
commutation relations with the Kac-Moody transformations $\delta$ and
$\td\delta$ defined in (\ref{deltacV})--(\ref{tddeltaX}) are given
by
\bea
{[}\delta_1^V, \delta_2{]} \!\!\!\ &=& \!\!\!\
  \fft{\lambda_1\lambda_2}{(\lambda_1-\lambda_2)}\,
\dot\delta(\lambda_2,\ep_2) + 
 \fft{\lambda_1\lambda_2}{(\lambda_1-\lambda_2)^2}\,
\big( \delta(\lambda_2,\ep_2) -\delta(\lambda_1,\ep_2)\big)\,,
\label{Vdelta}\\
{[}\delta_1^V,\td\delta_2{]}\!\!\! &=& \!\!\!\ 
 \fft{\lambda_1\lambda_2^2}{(1-\lambda_1\lambda_2)}\,
  \dot{\td\delta}(\lambda_2,\ep_2) + 
  \fft{\lambda_1\lambda_2}{(1-\lambda_1\lambda_2)^2}\,
   \td\delta(\lambda_2,\ep_2) +
  \fft{\lambda_1\lambda_2}{(1-\lambda_1\lambda_2)^2}\,
  \delta(\lambda_1,\ep_2)\,.\label{Vtddelta}
\eea
(We have calculated these commutators acting on all the fields, namely
$\cV$, $X$ and $\rho$.)
Substituting the expansions (\ref{deltaexps}) and (\ref{virexp1}) into
these expressions, and equating the coefficients of each power of 
$\lambda_1$ and $\lambda_2$, we can read off the commutation relations
\bea
{[}\delta^V_\m, \delta_\n {]} &=& -n \delta_{\sst{(n+m)}}\,,
\qquad m\ge 1\,,\quad n\ge0\,,\label{com1}\\
{[}\delta^V_\m, \td\delta_\n {]} &=& n \td\delta_{\sst{(n-m)}} + 
n \delta_{\sst{(m-n)}}\,,\qquad m\ge 1\,,\quad n\ge1\,,\label{com2}
\eea
where in (\ref{com2}) it is understood that $\delta_{\sst{(m-n)}}$ is zero
if $n> m$ and that $\td\delta_{\sst{(n-m)}}$ is zero if $n\le m$.  The 
commutation relations (\ref{com1}) and (\ref{com2}) can be unified into
one formula if we defined the full set of Kac-Moody modes $\Delta_\n$
as in (\ref{Deltadef}).  We then find
\be
{[}\delta^V_\m, \Delta_\n {]} =  -n \Delta_{\sst{(n+m)}}\,,
\label{mixed1}
\ee
with $m\ge 1$ and $-\infty\le n\le \infty$.

   With our previous associations in which $\Delta_\n$ corresponds to
the Kac-Moody generators $J_n^i$, and 
$\delta^V_\m$ corresponds to the Virasoro generator $L_m$ with $m\ge 1$, 
we therefore have
\be
[L_m, J_n^i] = - n J_{n+m}^i\,.\label{VKM}
\ee
Note that it is because $\delta^V_\m$ is associated with the positive
half of the Virasoro algebra, and thus it can only {\it increase} the
Kac-Moody level number in (\ref{VKM}), that (\ref{Vdelta}) has only 
$\delta$ Kac-Moody transformations on the right-hand side, 
whereas (\ref{Vtddelta}) has both $\delta$ and $\td\delta$ Kac-Moody
transformations.

   We can also extend the Virasoro-like transformations from $\delta^V$
to $\hat\delta^V$ in these mixed commutator calculations.  The 
necessary calculations are just a simple extension of those already presented,
and the upshot is that one merely has to extend the range of the $m$
index in (\ref{mixed1}) and (\ref{VKM}) down to $m\ge -1$ when $\delta^V$
is replaced with $\hat\delta^V$.

\section{Conclusions}

   In this paper, we have extended our previous work on the symmetries of
two-dimensional symmetric-space sigma models, by now considering the case
where the sigma model is coupled to gravity.  These gravity-coupled sigma 
models arise in the toroidal reduction of gravity and supergravity
theories from higher dimensions.  There has been some considerable 
discussion of the Kac-Moody symmetries of these models in earlier literature,
and we took the
paper \cite{schwarz2} by Schwarz as the starting point for
our construction.

   We were able to improve on the construction of Kac-Moody transformations 
in \cite{schwarz2} in several important respects.  Firstly, we gave a
proper treatment of the use of the ``homogeneous'' Kac-Moody transformations
(\ref{deltahom}) to subtract out the pole in the ``inhomogeneous'' 
transformations (\ref{deltainhom}) that would otherwise occur at 
$\tau_1=\tau_2$.  The fact that this can be done simultaneously at all 
points $x^\mu$ in the two-dimensional spacetime depends upon the quite
remarkable identity (\ref{uvrel2}).  

   Secondly, were able to
obtain transformations for the full affine Kac-Moody extension $\hat\cG$ of
the manifest symmetry algebra $\cG$ of the coset $\cG/\cH$.  This result
extends the one obtained in \cite{schwarz2}, where only a certain subalgebra
$\hat\cG_H$ of symmetries was found.  

   We also constructed further infinite-dimensional symmetry 
transformations of the SSM fields that are singlets under the original
$\cG$ Lie algebra.  They had not been found in \cite{schwarz2}.  These
symmetries correspond to a subalgebra of the centreless Virasoro algebra,
corresponding to the generators $L_n$ with $n\ge -1$. 
It would be of considerable interest to see if there exist further 
symmetries that could extend this subalgebra to the full Virasoro algebra.

   Finally, we remark that all of our symmetry analyses have been restricted
purely to the infinitesimal level.  It would be very interesting to 
extend the methods used in this paper to the non-infinitesimal level.  This
would be of importance both from the perspective of understanding the
U-duality symmetries of string theory, and in order to make use of the
infinite-dimensional symmetries for the purpose of generating new
solutions from old ones.

\section*{Acknowledgements}

   We are very grateful to John Schwarz for discussions, and for
drawing our attention to references \cite{schwarz1,schwarz2}.  We thank
also Hermann Nicolai for discussions.  This research has
been generously supported by George Mitchell and the Mitchell Family
Foundation.  The research of
H.L. and C.N.P. is also supported in part by DOE grant DE-FG03-95ER40917.

\newpage

\appendix

\section{The Flat-Space Limit}\label{appsec}

\subsection{Decoupling of gravity}

   In a previous paper \cite{lupepo}, we studied the somewhat simpler
problem of symmetric-space coset models in a purely flat two-dimensional
spacetime.  In that case, there is no conformal function $\rho$, and the
 equations (\ref{eom}) and (\ref{cartmau}) become simply
\be
d{*A}=0\,,\qquad dA+ A\wedge A=0\,.
\ee
It is evident, therefore, that we can recover the flat-space limit by taking
\be
\rho=\hbox{constant}\,,
\ee
and furthermore this is consistent with its equation of motion (\ref{rhoeom}).
In fact since in general $\rho$ has the solution $\rho=\rho_+(x^+) +
 \rho_-(x^-)$, as in (\ref{rhosol}), we can conveniently parameterise a
family of flat-space limits by 
\be
\rho_+ = 1+\gamma\,,\qquad \rho_- = 1-\gamma\,,\label{alphapar}
\ee
where $\gamma$ is a constant.

   In the flat-space limit the spectral function $\tau$ appearing the the
Lax equation (\ref{Lax2}) is a constant, which was denoted by $t$ in 
\cite{lupepo}.   From (\ref{tautheta}) and (\ref{thetasol2}), we see
that (\ref{alphapar}) implies
\be
\lambda= \fft{t}{1+2\gamma\, t + t^2}\,.\label{lamt}
\ee
Thus the constant spectral parameter $t$ used in \cite{lupepo} and the
constant spectral parameter $\lambda$ that we have been using in this paper
are not identical in the flat-space limit.  As a consequence, the
expansions of the Kac-Moody and Virasoro transformations in mode sums
as in (\ref{deltaexps}) and (\ref{virmodes}) will take different forms
depending on whether we use the $t$ parameter or the $\lambda$ parameter
in the flat-space limit, and this amounts to a change of basis for the
algebras.

   In what follows, we shall illustrate how the bases for the Kac-Moody and 
Virasoro algebras are related in the ``symmetric'' choice for the flat-space
limit, where we set $\gamma=0$ (and hence $\rho_+=\rho_-=1$).  The calculation
for a general choice of $\gamma$ proceeds in a very similar manner.  With
the choice $\gamma=0$ we have from (\ref{lamt}) that
\be
\lambda=\fft{t}{1+t^2}\,,\qquad t= \fft{1-\sqrt{1-4\lambda^2}}{2\lambda}\,.
\label{lamt2}
\ee
(We choose the negative root in the solution for $t$ so that small $\lambda$
corresponds to small $t$.)

\subsection{Kac-Moody symmetries in the flat-space limit}

   First, let us consider the Kac-Moody transformations $\delta_1 X_2$
defined in (\ref{deltaX}).  Using (\ref{uvsol}) and (\ref{uvrel2}), we
can first rewrite $\delta_1 X_2$ as
\be
\delta_1 X_2 =\fft{s_1}{\lambda_1\rho}\,\Big[ \fft{\tau_2}{(\tau_1-\tau_2)}\,
    (\eta_1 X_2-X_2\ep_1 X_2^{-1}) + \fft{\tau_1\tau_2}{(1-\tau_1\tau_2)}\,
(M^{-1}\eta_1^\dagger M + X_2\ep_1X_2^{-1})\Big] X_2\,.
\ee
Taking the flat-space limit with $\rho=2$, $\tau_i=t_i$ and $\lambda_i$
related to $t_i$ by (\ref{lamt2}), we therefore have
\bea
\delta_1^{\rm lim} X_2 &=& \fft{1+t_1^2}{1-t_1^2}\, \Big[
  \fft{t_2}{(t_1-t_2)}\, (\eta_1 X_2-X_2\ep_1 X_2^{-1}) +
         \fft{t_1 t_2}{(1-t_1 t_2)}\, M^{-1}\eta_1^\dagger M\Big] X_2\nn\\
 && +
 \fft{1+t_1^2}{1-t_1^2}\, \fft{t_1 t_2}{(1-t_1 t_2)}\, X_2\ep_1\,.
\eea
(The superscript ``lim'' denotes the flat-space limit of the general
gravitationally-coupled transformations we have derived in this paper.)
Comparing with the definitions of the Kac-Moody transformations that
we used in the purely flat-space discussion in \cite{lupepo}, and which
we denote by $\delta_1^{\rm flat}$ and $\td\delta_1^{\rm flat}$ for
this present discussion, we read off that 
\be
\delta_1^{\rm lim} X_2= \fft{1+t_1^2}{1-t_1^2}\,(\delta_1^{\rm flat} 
            +\td\delta_1^{\rm flat}) X_2\,.\label{oldnew1}
\ee

   By definition, the modes of $\delta_1^{\rm lim}$ are read off from 
an expansion in powers of $\lambda_1$, as in (\ref{deltaexps}).  Also,
by definition, the modes of $\delta_1^{\rm flat}$ and 
$\td\delta_1^{\rm flat}$ that we used in the flat-space situation in 
\cite{lupepo}) were expanded in powers of $t_1$.  Thus we have only to
substitute the expansions into (\ref{oldnew1}) to obtain
\be
\sum_{n\ge0}\lambda_1^n \delta_\n^{\rm lim} = \fft{1+t_1^2}{1-t_1^2}\,
\Big(\delta_{\sst{(0)}}^{\rm flat} + 
 \sum_{n\ge 1} t_1^n (\delta_\n^{\rm flat} + \td\delta_\n^{\rm flat})
      \Big)\,.
\ee
We now use (\ref{lamt2}) to express $t_1$ in terms of $\lambda_1$ on the
right-hand side, and equate coefficients of each power of $\lambda_1$.  It
is convenient to recall from \cite{lupepo} that, just as we did for the
the gravity-coupled case in this paper in (\ref{Deltadef}), the full set 
of Kac-Moody 
transformations $\Delta_\n^{\rm flat}$ were defined from the non-negative
modes $\delta_\n^{\rm flat}$  and the negative modes $\td\delta_\n^{\rm flat}$
by $\Delta_\n^{\rm flat}= \delta_\n^{\rm flat}$ ($n\ge0$) and 
$\Delta_\n^{\rm flat}= \td\delta_{\sst{(-n)}}^{\rm flat}$ ($n\le -1$).  
We then find that
\bea
\delta^{\rm lim}_{\sst{(0)}} &=& \Delta_{\sst{(0)}}^{\rm flat}\,,\quad
\delta^{\rm lim}_{\sst{(1)}} = \Delta_{\sst{(-1)}}^{\rm flat} +
                     \Delta_{\sst{(1)}}^{\rm flat}\,,\quad
\delta^{\rm lim}_{\sst{(2)}} = \Delta_{\sst{(-2)}}^{\rm flat} +
                          2 \Delta_{\sst{(0)}}^{\rm flat}+
                     \Delta_{\sst{(2)}}^{\rm flat}\,,\nn\\
\delta^{\rm lim}_{\sst{(3)}} &=& \Delta_{\sst{(-3)}}^{\rm flat} +
                          3 \Delta_{\sst{(-1)}}^{\rm flat}+
                         3 \Delta_{\sst{(1)}}^{\rm flat}+
                     \Delta_{\sst{(3)}}^{\rm flat}\,,
\eea
and so on, with the general $n$ case given by
\be
\delta^{\rm lim}_\n = \sum_{p=0}^n C^n_p\, \Delta_{\sst{(n-2p)}}^{\rm flat}\,,
\qquad C^n_p\equiv \fft{n!}{p!\, (n-p)!}\,.
\ee
This shows that the non-negative half of the Kac-Moody algebra in the
flat-space limit of the gravity-coupled models is related to a combination
of the positive and negative halves of the Kac-Moody algebra that arose in 
the previous flat-space discussion in \cite{lupepo}. 

   We turn now to the negative half of the Kac-Moody algebra of
the present paper, described by $\td\delta_1$ in (\ref{tddeltaX}).  We have
\be
\td\delta_1 X_2 = \fft{\lambda_1\lambda_2}{(1-\lambda_1\lambda_2)}\,
     X_2 \ep_1\,,
\ee
from which it follows that
\be
\td\delta_\n^{\rm lim} X_2 = \lambda_2^n\, X_2\ep_1\,.\label{lamv}
\ee
Equally, in the flat-space expansion from \cite{lupepo} we have
\be
\td\delta_\n^{\rm flat} X_2 = t_2^n\, X_2\ep_1\,.\label{tv}
\ee
By substituting (\ref{lamt2}) into (\ref{lamv}), expanding the 
right-hand side in powers of $t_2$, and then using (\ref{tv}), we can 
read off the expressions for the modes $\td\delta_\n^{\rm lim}$ in terms
of the modes $\td\delta_\n^{\rm flat}$.  The first few modes are 
given by
\bea
\td\delta_{\sst{(1)}}^{\rm lim} &=& \td\delta_{\sst{(1)}}^{\rm flat}
 -\td\delta_{\sst{(3)}}^{\rm flat} + \td\delta_{\sst{(5)}}^{\rm flat}
  -\td\delta_{\sst{(7)}}^{\rm flat}+\cdots\,,\nn\\
\td\delta_{\sst{(2)}}^{\rm lim} &=& \td\delta_{\sst{(2)}}^{\rm flat}
 -2\td\delta_{\sst{(4)}}^{\rm flat} + 3\td\delta_{\sst{(6)}}^{\rm flat}
  -4\td\delta_{\sst{(8)}}^{\rm flat}+\cdots\,,\nn\\
\td\delta_{\sst{(3)}}^{\rm lim} &=& \td\delta_{\sst{(3)}}^{\rm flat}
 -3\td\delta_{\sst{(5)}}^{\rm flat} + 6\td\delta_{\sst{(7)}}^{\rm flat}
  -10\td\delta_{\sst{(9)}}^{\rm flat}+\cdots\,,
\eea
and the general case is given by
\be
\td\delta_\n^{\rm lim}= \sum_{p=0}^\infty \fft{(-1)^p\,(n+p-1)!}{p!\, (n-1)!}\,
       \td\delta_{\sst{(n+2p)}}^{\rm flat}\,,\qquad n\ge 1\,.
\ee

    Combining our results for $\delta_\n^{\rm lim}$ and 
$\td\delta_\n^{\rm lim}$, and using the definition (\ref{Deltadef}) for
the full set of Kac-Moody transformations, we therefore have
\bea
\Delta_\n^{\rm lim} &=& \sum_{p=0}^n C^n_p\, 
\Delta^{\rm flat}_{\sst{(n-2p)}}\,,\qquad
   n\ge 0\label{dela}\\
\Delta_\n^{\rm lim} &=& \sum_{p=0}^n (-1)^p\, C^{p-n-1}_p\, 
   \Delta^{\rm flat}_{\sst{(n-2p)}}\,,\label{delb}
\qquad  n\le -1\,.
\eea
Now if $n$ is temporarily generalised to be an arbitrary real variable 
whilst $p$ is an
integer, then we have $C^n_p/C^{p-n-1}_p =(-1)^p$, and so we see 
that (\ref{dela})
and (\ref{delb}) can really be combined into a single formula of the 
form (\ref{dela}), with the $p$ summation appropriately extended;
\be
\Delta_\n^{\rm lim} = \sum_{p=0}^\infty C^n_p\,
\Delta^{\rm flat}_{\sst{(n-2p)}}\,,\qquad n\in\Z\,.
\ee

   A
direct
demonstration that $\Delta_\n^{\rm lim}$ satisfies the standard Kac-Moody
algebra (\ref{kacmoody}) if $\Delta_\n^{\rm flat}$ satisfies the 
standard Kac-Moody algebra
can easily be given, based on the identity that 
$(1+t)^p (1+t)^q = (1+t)^{p+q}$.

\subsection{Virasoro-type symmetries in the flat-space limit}

    In our discussion of the Virasoro-type symmetries in section 
\ref{virasorosec}, we found that not only the fields $\cV$ and $X$, but
also the field $\rho$, is subject to these transformations, 
as in (\ref{finalV}).  When we take the flat-space limit we are fixing 
$\rho$ to a constant, and consequently some of the Virasoro-type symmetry
of the general gravity-coupled case will be broken.  In fact, as we shall
see below, it is the two ``extra'' generators $\hat\delta^V_{\sst{(0)}}$ and 
$\hat\delta^V_{\sst{(-1)}}$, whose action is given in (\ref{virextra}), that
are broken when we fix $\rho$ (and also $\rho_+$ and $\rho_-$ separately).  
In fact these two transformations will be treated as compensating 
transformations that restore $\rho_+$ and $\rho_-$ to their chosen fixed
values when the remaining Virasoro-type transformations  
$\hat\delta^V_\n$ with $n\ge1$ act.  This loss of the $n=0$ and $n=-1$ 
symmetries in the flat-space limit is consistent with the fact that they
were never seen in the purely flat-space results in \cite{lupepo}.

   It can be seen from (\ref{deltarhoprhom}) (with $\beta=0$) and  
(\ref{virextra}) that in order to preserve the flat-space limit with
$\rho_+=\rho_-=1$, the $\delta_1^V$ Virasoro transformations for the
modes $n\ge1$ should be accompanied by $\hat\delta^V_{\sst{(0)}}$ and
$\hat\delta^V_{\sst{(-1)}}$ compensating transformations in the combination
\be
\td\delta_1^V = \delta_1^V - s_1^2\, \hat\delta^V_{\sst{(0)}} -
                     2 s_1 c_1\, \hat\delta^V_{\sst{(-1)}}\,.
\label{compvir}
\ee
We can therefore now pass to the flat-space limit by setting 
$\rho_+=\rho_-=1$, and calculate the commutators of these
compensated Virasoro-type transformations, using the expressions for the
commutators of the transformations appearing on the right-hand side of
(\ref{compvir}) that we have obtained in section \ref{virasorosec}.  Thus
we find that
\bea
{[} \td\delta_1^V,\td\delta_2^V {]} &=& 
\fft{\lambda_1\lambda_2}{\lambda_1-\lambda_2}\, \left[
\fft{1-4\lambda_1^2}{1-4\lambda_2^2}\, \del_{\lambda_1} \td\delta_1^V +
 \fft{1-4\lambda_2^2}{1-4\lambda_1^2}\, \del_{\lambda_2} 
  \td\delta_2^V\right]\nn\\
&&- \fft{2\lambda_1\lambda_2}{(\lambda_1-\lambda_2)^2}\,
\left[ \fft{1+4\lambda_1^2-8\lambda_1\lambda_2}{1-4\lambda_2^2}\, 
    \td\delta_1^V -
   \fft{1+4\lambda_2^2-8\lambda_1\lambda_2}{1-4\lambda_1^2}\,
    \td\delta_2^V\right]\,.\label{tdVcom}
\eea

   Since $\cV$ is inert under $\delta_{\sst{(0)}}^V$ and 
  $\delta_{\sst{(-1)}}^V$, it follows that $\td\delta_1^V\cV=\delta_1^V \cV$,
and so from (\ref{deltaVcV}) and (\ref{hsol2}) we have, after taking the
flat-space limit in which $\lambda$ and $t=\tau$ are related by (\ref{lamt2}),
\bea
\td\delta^V \cV &=& -\fft{t}{1-t^2}\, \cV \big( \del_{\lambda} X\big) X^{-1}
                            \,,\nn\\
&=& -\fft{t}{1-t^2}\, \fft{\del t}{\del\lambda}\, 
             \big( \del_t X\big) X^{-1}\,,\nn\\
  &=& \left(\fft{1+t^2}{1-t^2}\right)^2\, \delta^{V,\, {\rm flat}}\, \cV\,,
\label{virlimflat}
\eea
where $\delta^{V,\, {\rm flat}}$ is the Virasoro-type transformation that
we found in the purely flat-space discussion in \cite{lupepo}.  (It is
given by $\delta^{V,\, {\rm flat}} \cV = -t \big(\del_t X(t)\big) X(t)^{-1}$.)
We have also verified that the same relation (\ref{virlimflat}) holds
when acting on $X$ instead of $\cV$.  
Using (\ref{virlimflat}), we can translate the commutator (\ref{tdVcom}) into
the commutator ${[} \delta^{V,\, {\rm flat}}_1, \delta^{V,\, {\rm flat}}_2 
 {]}$, finding
\bea
{[} \delta^{V,\, {\rm flat}}_1, \delta^{V,\, {\rm flat}}_2 {]} &=&
-2 t_1 t_2 \left[ \fft{1}{(t_1-t_2)^2} +\fft1{(1-t_1 t_2)^2}\right]
  \, \delta_1^{V,\,{\rm flat}}
  +\fft{t_1t_2(1-t_1^2)}{(t_1-t_2)(1-t_1 t_2)} \, \del_{t_1}
\delta_1^{V\, {\rm flat}}\nn\\
&&
-[1\leftrightarrow 2]\,,\label{comVM}
\eea
This agrees precisely with the commutator of Virasoro-type transformations 
that we found previously for the flat space models in \cite{lupepo}.

   The relation (\ref{virlimflat}) can be re-expressed as a relation
between the modes in the expansions of the two variations, namely
\be
\td\delta_1^V =\sum_{n\ge1} \lambda_1^n\, \td\delta_\n^V\,,\qquad
\delta_1^{V\,{\rm flat}} = \sum_{n\ge 1} t_1^n\,
     \delta_\n^{V\,{\rm flat}}\,.
\ee
For the first few levels we find
\bea
 \td\delta_{\sst{(1)}}^V &=& \delta_{\sst{(1)}}^{V\,{\rm flat}}\,,\qquad
 \td\delta_{\sst{(2)}}^V = \delta_{\sst{(2)}}^{V\,{\rm flat}}\,,\qquad
\td\delta_{\sst{(3)}}^V = \delta_{\sst{(3)}}^{V\,{\rm flat}} +
                      5 \delta_{\sst{(1)}}^{V\,{\rm flat}}\,,\nn\\
\td\delta_{\sst{(4)}}^V &=& \delta_{\sst{(4)}}^{V\,{\rm flat}} +
                      6 \delta_{\sst{(2)}}^{V\,{\rm flat}}\,,\qquad
\td\delta_{\sst{(5)}}^V = \delta_{\sst{(5)}}^{V\,{\rm flat}} +
                      7 \delta_{\sst{(3)}}^{V\,{\rm flat}}+
                     22 \delta_{\sst{(1)}}^{V\,{\rm flat}}\,.
\eea

\end{document}